\documentstyle[11pt,paspconf,epsfig]{article}

\begin{document}

\begin{flushright} FISIST/14-99/CENTRA\\
August 31, 1999
\end{flushright}

\title {PSCz-1.2 Jy Comparison: A Spherical Harmonics Approach}
\author {Lu\'{\i}s  Teodoro$^1$$^3$}
\affil{$^1$Instituto Superior T\'{e}cnico, Lisboa, Portugal}
\author{Enzo Branchini$^2$, Carlos Frenk$^3$, 
Will Saunders$^4$, Seb Oliver$^5$, Oliver Keeble$^5$, 
Michael Rowan-Robinson$^5$, Steve Maddox$^6$,
George Efstathiou$^6$, Will Sutherland$^7$, Simon White$^8$}
\affil{$^2$University of Groningen, Groningen, The Netherlands \newline
$^3$University of Durham, Durham, UK \newline
$^4$Edinburgh University, Edinburgh, UK\newline
$^5$Imperial College, London, UK\newline
$^6$University of Cambridge, Cambridge, UK\newline
$^7$University of Oxford, Oxford, UK\newline
$^8$Max Planck Institute for Astrophysics, Munich, Germany}

\begin{abstract} 
We perform a detailed comparison of the IRAS PSCz~and 1.2-Jy spherical
harmonic coefficients of the density and velocity fields in redshift
space. The monopole terms predicted from the two surveys show some
differences.  The mismatch between the velocity monopoles arises from
faint sources and disappears when extracting a PSCz subsample of
galaxies with fluxes larger than 1.2 Jy. The analysis of PSCz~dipole
moments confirms the same inconsistencies found by Davis, Nusser and
Willick (1996)\nocite{Davis:1996} between the IRAS 1.2-Jy gravity
field and {\it MARK}~III~peculiar velocities. We conclude that
shot-noise, which is greatly reduced in our PSCz~gravity field, cannot
be responsible for the observed mismatch.
\end{abstract}
\section{Introduction}
Nusser \& Davis (1994)\nocite{Nusser:1994} show that in linear
gravitational instability (GI) theory the peculiar velocity field in
redshift space is irrotational and thus can be expressed in terms of a
potential: ${\vec v}=-\nabla \Phi(\vec s)$. The angular
dependencies of the potential velocity field and the galaxy
overdensity field [both measured in redshift space and expanded in
spherical harmonics, $\Phi_{lm}(s)$ and $\hat{\delta}_{lm}(s)$,
respectively] are related by a modified Poisson equation:
\begin{equation}
{{1}\over{s^2}}{{d}\over{ds}}
\left( s^2 {{d\Phi_{lm}}\over{ds}} \right)
-{{1}\over{1+\beta}}
{{l(l+1)\Phi_{lm}}\over{s^2}}=
{{\beta}\over{1+\beta}}
\left( \hat{\delta}_{lm}-
{{1}\over{s}}{{d\ln{\phi}}\over{d \ln{s}}}
{{d \Phi_{lm}}\over{ds}} \right),
\label{eq:ndchapter5}
\end{equation}
where $\phi(s)$ is the selection function. To solve this differential
equation we first compute the density field on an angular grid
using cells of equal solid angle and
52 bins in redshift out to $s=18\,000$~kms$^{-1}$.
\begin{equation}
1+{\hat \delta}_j({\vec s}_n)=\frac{1}{(2\pi)^{3/2}\sigma_{1.2 n}^3}\sum_i^{N_j} \frac{1}{\phi(s_i)}\exp\left [
  -\frac{({\vec s}_n-{\vec s}_i)^2}{2\sigma_{1.2 n}^2} \right ]
\label{eq:smoothing5}
\end{equation}
where the sum is over all the galaxies within the catalogue $j$,
$N_j$.  The Gaussian smoothing width for the cell $n$ at redshift
$s_n$, $\sigma_{1.2 n}$, is given by $\sigma_{1.2 n} =
[\bar{n}_{1.2}\phi_{1.2}(s_n)]^{-1/3}$ (or 100~km s$^{-1}~$when such a
length is smaller than this), where $\bar{n}_{1.2}$~and
$\phi_{1.2}$~are the 1.2 Jy mean number density and selection
function, respectively.

\section{Datasets}
\label{section:datasets}
PSCz~is a new redshift survey which resulted from a
collaborative effort involving several British institutions (Durham,
Oxford, London, Edinburgh and Cambridge). The catalogue contains some
15500 IRAS PSC (Point Source Catalogue) galaxies with a 60
$\mu$m flux larger than 0.6-Jy. This survey covers 84$\%$ of the
sky. A more detailed description of the catalogue is given by Saunders
{\it et al.} 1999 (these proceedings). The 1.2-Jy catalogue (Fisher
{\it et al.}~1995\nocite{Fisher:1995}) contains 5321 IRAS~PSC
galaxies with a 60 $\mu m$ limit of 1.2-Jy which covers 87.6$\%$ of
the sky.

\section{Results, Discussion and Conclusions}
\label{section:conclusions5}
\begin{figure}
\begin{center}
\scalebox{0.55}{\includegraphics[200,250][415,685]{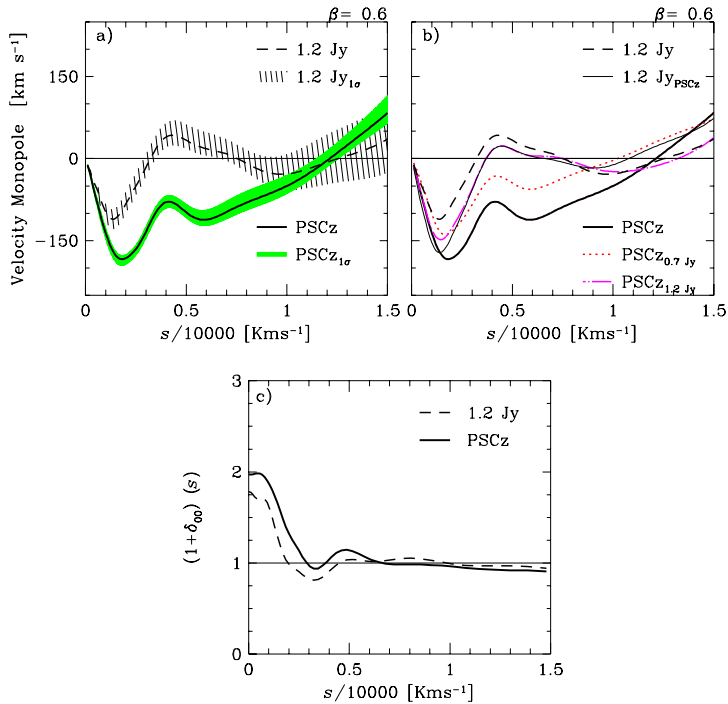}}
\caption{Monopole coefficients, $u_{00}(s)$ and ${\hat \delta}_{00}(s)$, 
 estimated from PSCz~and 1.2-Jy catalogues for $\beta =0.6$. On the
 two top panels dashed (thick) lines represent 1.2-Jy (PSCz)
 monopole mode of the velocity field.  a) The hatched and shaded
 regions indicate the 1.2 Jy and PSCz~1--$\sigma$~shot-noise
 uncertainties, respectively. b) Dotted, dot-dashed and thin lines
 represent PSCz$_{0.7}$, PSCz$_{1.2}$ and
 1.2-Jy$_{\mbox{PSCz}}$, respectively.  c) The thick line
 indicates the  PSCz~density monopole while the 1.2-Jy
 counterpart is represented by the dashed line.}
\label{fig:irasmonopole}
\end{center}
\end{figure}

We compare the line of sight peculiar velocity, $u ({\vec s})\equiv [ {\vec
v}({\vec v}) -{\vec v}_{LG} ] \cdot {\hat s}$, and density perturbations,
$\hat{\delta}(s)$, fields inferred from the IRAS PSCz~and 1.2-Jy
redshift surveys, in redshift-space.  To compute the spherical
harmonics coefficients up to $l_{max}=16$ we apply the algorithm
introduced by Nusser \& Davis (1994)\nocite{Nusser:1994} setting
$\beta=0.6$.

\begin{figure}
\begin{center}
\scalebox{0.55}{\includegraphics[30,260][565,685]{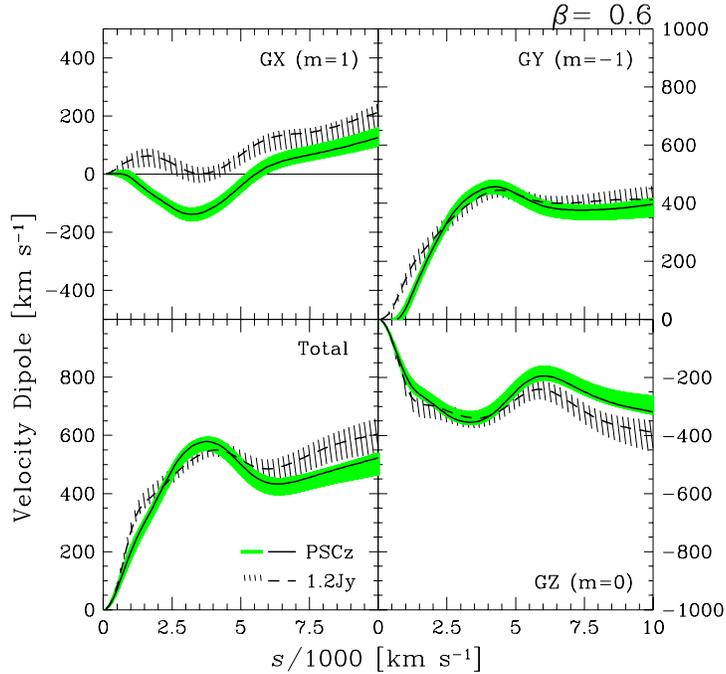}}
\caption{ The dipole coefficients, $u_{1m}(s)$\{$m=-1,0,1$\}, inferred from PSCz~and 1.2-Jy
for $\beta =0.6$. The solid and dashed lines in the various panels
represent PSCz~and 1.2-Jy dipoles, respectively. The $GX$, $GY$,
$GZ$~panels show the three Galactic components of the dipole and the
bottom left is their quadrature sum. Hatched and shaded regions
indicate the 1.2-Jy and PSCz~1--$\sigma$~shot-noise uncertainty,
respectively.}
\label{fig:irasdipoles}
\end{center}
\end{figure}

In Fig.~\ref{fig:irasmonopole} we display the monopole of the velocity
field, $u_{00}$ (two top panels) and the monopole of the density
${\hat \delta}_{00}$ (bottom panel). In the top-left panel, the
estimate of the velocity monopole in the 1.2-Jy (dashed line) is
systematically larger than the PSCz one (continuous thick line) within
a redshift of $s\approx 8000$~km s$^{-1}$.  The density monopole of
the two surveys show a similar qualitative radial dependency (bottom
panel). In Fig.~\ref{fig:irasdipoles} we show the three velocity
dipole terms of the two surveys along with the total amplitude (bottom
left panel).  The various dipole moments exhibit good agreement,
except $u_{11}$ (top left panel) for which there is a discrepancy of
100~km$s^{-1}$~in the redshift range $1500 - 7500$~kms$^{-1}$. Good
agreement is also found when comparing higher order multipoles
(Teodoro {\it et al.}~1999).  Note that when all the multipoles are
considered ({\it{i.e.}}~when performing a full v-v comparison) the
PSCz and 1.2-Jy gravity fields look fully consistent (Branchini {\it
et al.}~1999)\nocite{Branchini:1999}.

In all plots, shaded and hatched regions represent 1-$\sigma$
shot-noise uncertainties computed from 20 bootstrap realizations of
PSCz and 1.2 Jy surveys, respectively. Where does the discrepancy
between the monopoles of the PSCz and 1.2 Jy surveys come from? As
shown in the plots, the difference is larger than that expected from the
shot-noise. Tadros {\it et al.}~(1999)\nocite{Tadros:1999} have
suggested that the PSCz~catalogue may be incomplete for fluxes $\le
0.7$~Jy. If true, then we would expect that the velocity monopole for
the PSCz~with a flux cut at 0.7 Jy (PSCz$_{0.7}$) would be in good
agreement with the 1.2-Jy survey.  The dotted line
($u_{00,\mbox{PSCz}_{0.7}}$) in the top panel of
Fig. \ref{fig:irasmonopole} shows however that this is not case. It is only
in cutting the PSCz catalogue at a flux level of 1.2 Jy that, as
expected, the discrepancy disappears, provided that we use the same
mask for both catalogues. This is clearly seen in the top right panel
of Fig. \ref{fig:irasmonopole} in which the thin and dot-dashed lines
indicate 1.2-Jy$_{PSCz}$(1.2 Jy with the same mask as PSCz) and
PSCz$_{1.2-Jy}$ (PSCz with a 1.2 Jy flux limit) velocity monopoles,
respectively.

A more detailed comparison between the two catalogues is in progress in
which we assess the importance of systematic errors by using a suite of 
1.2 JY and PSCz mock catalogues drawn from N-body simulations.

\acknowledgments We thank Marc Davis for providing the original version 
of the reconstruction code and Adi Nusser for many dicussions.
LT has been supported by the grants PRAXIS XXI/BPD/16354/98 and 
PRAXIS/C/FIS/
13196/1998.

\end{document}